\documentclass{nature}

\title{The formation of dusty cold gas filaments from galaxy cluster simulations}

\author{Yu Qiu\hskip2pt\href{https://orcid.org/0000-0002-6164-8463}{\includegraphics[width=9pt]{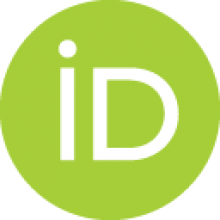}}$^{1,2*}$, Tamara Bogdanovi\'c\hskip2pt\href{https://orcid.org/0000-0002-7835-7814}{\includegraphics[width=9pt]{Orcid-ID.png}}$^2$, Yuan Li$^3$, Michael McDonald$^4$ \& Brian R. McNamara$^{5}$}

\usepackage{graphicx}
\makeatletter
\let\saved@includegraphics\includegraphics
\AtBeginDocument{\let\includegraphics\saved@includegraphics}
\renewenvironment*{figure}{\@float{figure}}{\end@float}
\makeatother

\usepackage[colorlinks=true,linkcolor=blue,anchorcolor=blue,citecolor=blue,filecolor=blue,menucolor=blue,runcolor=blue,urlcolor=blue]{hyperref}

\usepackage{color}
\usepackage{amsmath,amssymb,url}
\usepackage[normalem]{ulem}

\usepackage{lineno}
\usepackage{etoolbox} 
\makeatletter 
\newcommand*\linenomathpatch{\@ifstar{\linenomathpatch@AMS}{\linenomathpatch@}}
\newcommand*\linenomathpatch@[1]{
  \expandafter\pretocmd\csname #1\endcsname {\linenomathWithnumbers}{}{}
  \expandafter\pretocmd\csname #1*\endcsname{\linenomathWithnumbers}{}{}
  \expandafter\apptocmd\csname end#1\endcsname {\endlinenomath}{}{}
  \expandafter\apptocmd\csname end#1*\endcsname{\endlinenomath}{}{}
}
\newcommand*\linenomathpatch@AMS[1]{
  \expandafter\pretocmd\csname #1\endcsname {\linenomathWithnumbersAMS}{}{}
  \expandafter\pretocmd\csname #1*\endcsname{\linenomathWithnumbersAMS}{}{}
  \expandafter\apptocmd\csname end#1\endcsname {\endlinenomath}{}{}
  \expandafter\apptocmd\csname end#1*\endcsname{\endlinenomath}{}{}
}
\let\linenomathWithnumbersAMS\linenomathWithnumbers
\patchcmd\linenomathWithnumbersAMS{\advance\postdisplaypenalty\linenopenalty}{}{}{}
\makeatother 

\linenomathpatch{equation}
\linenomathpatch*{gather}
\linenomathpatch*{multline}
\linenomathpatch*{align}
\linenomathpatch*{alignat}
\linenomathpatch*{flalign}

\definecolor{Q-color}{rgb}{0.5,0.1,0.5}

\usepackage[export]{adjustbox}

\begin{document}

\spacing{1}

\maketitle

\begin{affiliations}
\small 
 \item Kavli Institute for Astronomy and Astrophysics, Peking University, Beijing, China
 \item Center for Relativistic Astrophysics, Georgia Institute of Technology, Atlanta, GA, USA
 \item Department of Astronomy, University of California, Berkeley, CA, USA
 \item Kavli Institute for Astrophysics and Space Research, Massachusetts Institute of Technology, Cambridge, MA, USA
 \item Department of Physics and Astronomy, Waterloo Centre for Astrophysics, University of Waterloo, Waterloo, ON, Canada
\end{affiliations}


\spacing{1}
\begin{abstract}
Galaxy clusters are the most massive collapsed structures in the universe whose potential wells are filled with hot, X-ray emitting intracluster medium. Observations however show that a significant number of clusters (the so-called cool-core clusters) also contain large amounts of cold gas in their centres, some of which is in the form of spatially extended filaments spanning scales of tens of kiloparsecs\cite{Hu1985, Edge2001}. These findings have raised questions about the origin of the cold gas, as well as its relationship with the central active galactic nucleus (AGN), whose feedback has been established as a ubiquitous feature in such galaxy clusters\cite{Fabian2012, Voit2015a, McNamara2016}.
Here we report a radiation hydrodynamic simulation of AGN feedback in a galaxy cluster, in which cold filaments form from the warm, AGN-driven outflows with temperatures between $10^4$ and $10^7$\,K as they rise in the cluster core. Our analysis reveals a new mechanism, which, through the combination of radiative cooling and ram pressure, naturally promotes outflows whose cooling time is shorter than their rising time, giving birth to spatially extended cold gas filaments.
Our results strongly suggest that the formation of cold gas and AGN feedback in galaxy clusters are inextricably linked and shed light on how AGN feedback couples to the intracluster medium.
\end{abstract}


Existing models for the formation of cold gas in galaxy clusters predict that filaments form in situ, out of thermally unstable intracluster medium (ICM) and flow toward the cluster core\cite{Gaspari2012a, Li2015, Prasad2015}. These models face several challenges because (a) they predict low values of the cooling-to-freefall timescale ratio ($t_{\rm cool}/t_{\rm ff}<10$), at odds with observations\cite{McNamara2016, Hogan2017}, and (b) because the filaments form from the ambient ICM with temperatures $\sim {\rm few} \times 10^7\,$K, they are expected to be devoid of dust, in contrast with observations\cite{Sparks1989, Goudfrooij1994, Donahue2011}. In this work we present a different picture, in which the cold filaments form from the warm phase of the multiphase outflows driven by AGN feedback, while they are rising. {Because of the moderate temperature of the warm outflows, filaments that form in this way can preserve the dust content drawn from the cluster core (Methods)}. They also have spatial and velocity distributions consistent with the H$\alpha$ filaments seen in Perseus and other cool-core clusters\cite{Conselice2001b, Fabian2008, McDonald2010, GM2018, Qiu2019}. In addition, the properties of the X-ray emitting plasma in our simulations are in good agreement with those observed in clusters and early-type galaxies\cite{Hogan2017, Babyk2018a}, with minimum cooling-to-freefall timescale ratios around 20, as required in a self-consistent picture of the ICM evolution.

\spacing{1}
\begin{figure}[t!]
\centering
\includegraphics[width=\linewidth]{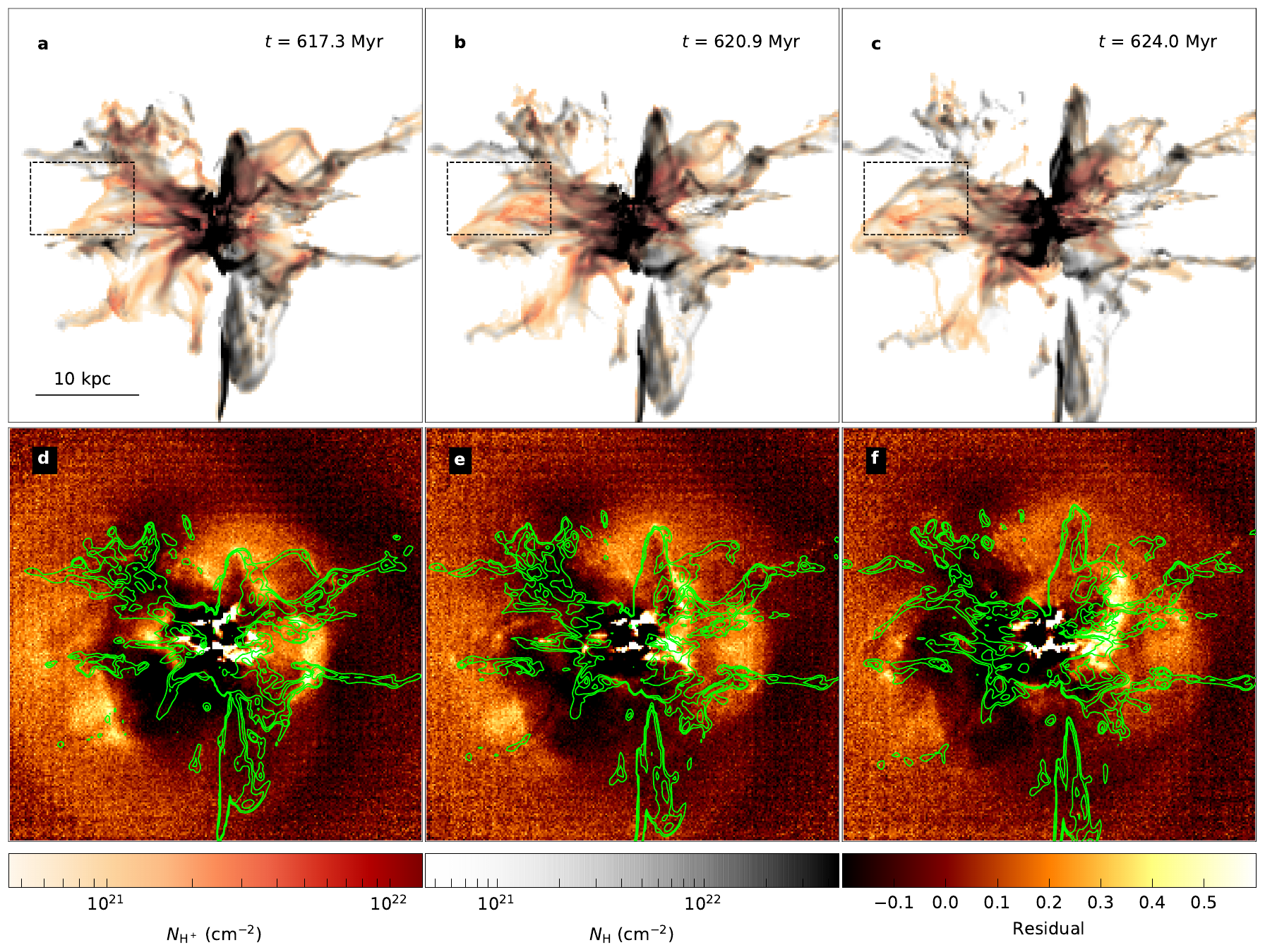}
\caption{\small {\bf Formation of cold gas filaments from warm galactic outflows.} 
{\bf a-c}, Outflows launched from the cluster core are dominated by warm, ionized gas ($10^4\,\rm{K}< T \leq 10^7$\,K, orange), whose cooling timescale is shorter than the rising timescale, resulting in formation of a spatially extended network of cold, neutral filaments ($T\leq 10^4\,$K, grey). Colour intensity represents the column density of ionized / neutral hydrogen. The dashed box highlights a region where the cold filament forms on the edge of the warm outflow over $\sim7\,$Myr as it rises to about 20\,kpc from the cluster centre. The AGN jet axis is fixed along the vertical direction throughout. A movie spanning 20\,Myr of evolution is provided in Supplementary Video 1. 
{\bf d-f}: Fractional variance of the X-ray photon count of the cluster core in the $1-7$\,keV band, in snapshots corresponding to {\bf a-c}. Green contours trace the cold gas with column density $N_{\rm H}=(3,\ 6,\ 12)\times10^{21}\,{\rm cm}^{-2}$. Features in the inner 3\,kpc are an artifact of the visualization, that arises due to large variation in the X-ray surface brightness in this region.
}
\label{fig:pic}
\end{figure}

In order to capture this process, we performed 3D radiation-hydrodynamic simulations of a cool-core cluster (CCC) with the code \texttt{Enzo}\cite{Bryan2014}. In this setup, the central AGN, powered by accretion onto the supermassive black hole (SMBH), is a source of both radiative and kinetic feedback. A finding that emerged from these simulations is that episodes of intense AGN feedback coincide with the appearance of spatially extended cold gas filaments in the cluster core\cite{Qiu2018}. In order to understand the relation between AGN feedback and formation of the cold gas, we re-simulated an episode featuring extended filaments with much higher temporal cadence of outputs, and use it to follow their evolution from formation to infall (Methods). 

Fig.~\ref{fig:pic}a-c shows the spatial distribution of the AGN-driven outflows in a sequence of snapshots from this simulation. Because the central SMBH resides in an environment that includes cold ($T\leq10^4$\,K), warm ($10^4<T\leq10^7$\,K) and hot ($T>10^7$\,K) gas, the central AGN may in principle launch multiphase outflows. We however find that the outflows which extend to $\sim20-30$\,kpc mostly comprise diffuse warm gas and a network of filamentary cold gas. The cold filaments in our simulation exhibit a spatial distribution consistent with the H$\alpha$-emitting filaments seen in Perseus\cite{Conselice2001b, Fabian2008, GM2018} and other cool-core clusters\cite{McDonald2010}. Fig.~\ref{fig:pic}d-f illustrates their relationship with the X-ray emitting ICM. These panels show the simulated X-ray images, calculated as the fractional variance from the azimuthally-averaged X-ray photon count (Methods). Dark areas correspond to regions with a deficit of the X-ray-emitting plasma and are similar to the X-ray cavities observed in many CCCs\cite{Birzan2004, McNamara2005}. These are a consequence of AGN feedback, which inflates low density bubbles and heats the surrounding ICM. Because the cold filaments and the X-ray cavities are both linked to AGN-driven outflows, the two phenomena are contemporaneous and co-spatial\cite{Russell2016a,McDonald2019}.

The cold filaments are therefore unambiguously associated with the outflows and the central goal of this work is to understand whether they (a) are entrained by the outflows and lifted from the core, where they existed in the first place, or (b) form from the warm outflows as they are rising. The former hypothesis suffers from a disadvantage that initially stationary or infalling cold gas clumps, which tend to be dense and have small cross-sections, are difficult to accelerate in outflows en masse. Even if such acceleration mechanism exists, the cold gas would need to be launched with speeds $\gtrsim1,000\,{\rm km\,s}^{-1}$ in order to reach the observed altitudes of tens of kpc. This is in conflict with observations that consistently find lower speeds for cold filamentary gas\cite{McDonald2012}. At the same time, cold gas would have to avoid instability and dissipation given the short cloud-shredding timescale\cite{Zhang2017}. The latter hypothesis is also more compelling because the warm gas phase has cooling timescale $t_{\rm cool}\lesssim10$\,Myr, considerably shorter than the rising timescale of outflows (comparable to freefall timescale $\sim$100\,Myr at 30\,kpc). Therefore, the cold filaments can form naturally from the warm outflows as the they are rising in the potential well of the cluster and our aim is to examine whether this hypothesis is supported by our high cadence simulation.


\begin{figure}[t!]
\centering
\includegraphics[width=\linewidth]{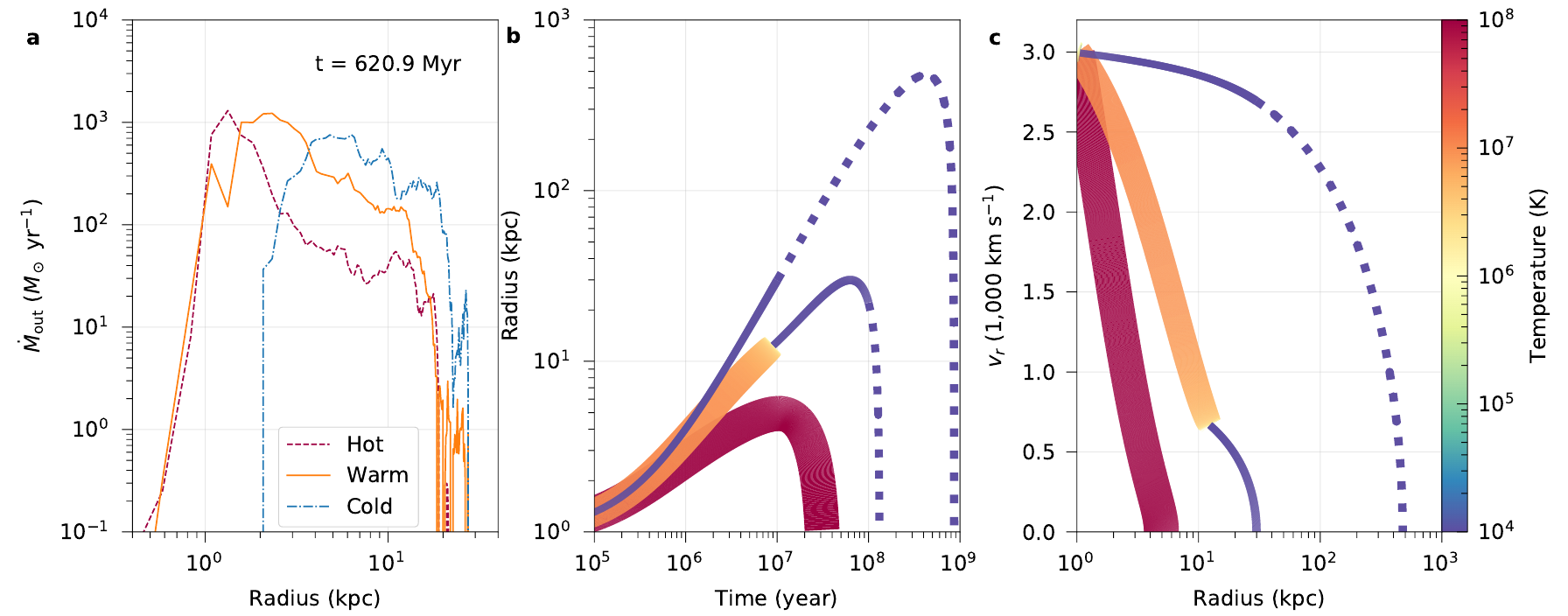}
\caption{\small {\bf Evolution of AGN-driven outflows for gas of different temperatures.} 
{\bf a}, Mass outflow rate as a function of radius for hot ($T>10^7$\,K; {red dashed}), warm ($10^4\,\rm{K}<T\leq10^7$\,K; {orange solid}), and cold ($T\leq 10^4$\,K; {blue dot-dashed}) gas with radial velocity $v_r>300\,{\rm km\,s}^{-1}$, corresponding to Fig.~\ref{fig:pic}b. Warm and hot gas dominate below 4\,kpc, while the cold gas dominates at larger radii. 
{\bf b}, Modeled trajectories of gas clumps with initial temperatures of $10^6$, $10^7$, and $10^8\,{\rm K}$, in pressure equilibrium with the intracluster medium. {The line colour represents the instantaneous temperature of the clump (see colour bar on the right). The width of each band is proportional to the logarithm of the clump size. The width of the middle band changes abruptly due to rapid ($<10^5$\,yr) radiative cooling of the gas from $10^{6}$ to $10^{4}$ \,K}. The ram pressure confines the $10^8\,{\rm K}$ (dilute) clumps to the central few kpc, but $10^6$\,K and $10^7$\,K (denser) clumps reach distances of $\sim 10-100\,{\rm kpc}$. The cold clumps are expected to dissipate in the ambient hot gas after $\sim10^{7-8}$ years, as indicated by by the dotted lines (see Methods).
{\bf c}, Radial velocity as a function of radius for the clumps shown in panel {\bf b}. The $10^6$\,K clump effectively follows a ballistic trajectory. In comparison, the trajectories of the $10^7$ and $10^8$\,K clumps differ due to the effect of ram pressure. 
}
\label{fig:model}
\end{figure}

To understand why cold filaments form preferentially as a result of the radiative cooling of the warm gas, we inspect the spatial distribution of different outflow phases. Fig.~\ref{fig:model}a shows the mass outflow rate of different components as a function of radius, $\dot{M}_{\rm out}(r)$. We calculate $\dot{M}_{\rm out}(r)$ by dividing the cluster into 0.25\,kpc-thick nested shells, and in each shell compute the volume average of $4\pi r^2 \rho v_r$, where $\rho$ is the density, and $v_r$ is the radial velocity of the outflowing gas. In order to focus on the outflows (as opposed to the gas that exhibits turbulent or other local motion), we only include $\dot{M}_{\rm out}$ for the gas with $v_r > 300\,{\rm km\,s}^{-1}$. Fig.~\ref{fig:model}a reveals that the three temperature components exhibit a degree of spatial stratification. Heated by the AGN radiative feedback, the hot outflows dominate in the inner 2\,kpc but their mass rate declines beyond this radius. The warm outflows dominate in the region between 2 and 4\,kpc, followed by the cold gas which dominates to a few tens of kpc.

We introduce a simple analytic model to elucidate the stratification of different outflow phases in our simulation. In Fig.~\ref{fig:model}b we show the trajectories of spherical gas clumps with initial temperatures $10^6$, $10^7$, $10^8$\,K and mass $10^8\,M_\odot$, in pressure equilibrium with the ICM. {In the illustrated scenario, the clumps are launched with the same fiducial speed of $3,000\,{\rm km\,s}^{-1}$ from the radius of 1\,kpc. This value corresponds to the maximum launching speed in the simulation, and the outflows launched at lower speeds will reach lower altitudes and produce filaments with lower line-of-sight velocity.} The temperature of each clump evolves primarily due to radiative cooling, and to a smaller degree due to the adiabatic expansion or compression in the ambient medium. The clumps also experience ram pressure from the ICM, which confines $10^8$\,K (dilute) clumps to the central few kpc. The effect of ram pressure on $10^6$\,K and $10^7$\,K clumps is considerably weaker. This is because their cooling timescales are shorter, causing the cross section of these clumps to decrease while they are rising. As a consequence, they can travel farther in distance. The ram pressure therefore naturally promotes outflows whose cooling time is shorter than their rising time and hinders outflows whose cooling time is longer.

Fig.~\ref{fig:model}c shows the radial velocity as a function of radius, $v_r (r)$, for the same clumps. In this panel, $10^6$\,K clump cools to $10^4$\,K within only $10^5$ yr after it is launched, and is largely unaffected by ram pressure after that point. {Although in this illustration we assume that $10^6$\,K gas is launched with the same velocity as the hotter gas for simplicity, this is unlikely to happen in reality because of a difficulty to accelerate it in such a short cooling timescale. We nevertheless show it in Fig.~\ref{fig:model}b,c as an illustration of a fiducial ballistic trajectory in contrast to the $10^7$ and $10^8$\,K clumps, whose trajectories are visibly affected by ram pressure.} In the case of the $10^7$\,K clump, radiative cooling results in the rapid evolution of its properties at $r\approx10$\,kpc. This alleviates the impact of ram pressure, and allows what is now cold and compact gas clump to travel beyond 20\,kpc. The $10^8$\,K clump does not undergo a similar transition and remains confined to the inner few kpc. It is worth noting that the radial velocity of the cold clumps, at the radii where they spend most of the time, is below $\sim$1,000\,km\,s$^{-1}$ (see also Methods). These values are consistent with the line-of-sight velocity distribution of the cold gas measured in observations of CCCs\cite{McDonald2012,GM2018}. Therefore, warm outflows with $T\lesssim10^7$\,K can naturally produce the extended cold gas filaments with low line-of-sight velocities\cite{Qiu2019}.


\begin{figure}[ht!]
\centering
\includegraphics[width=\linewidth]{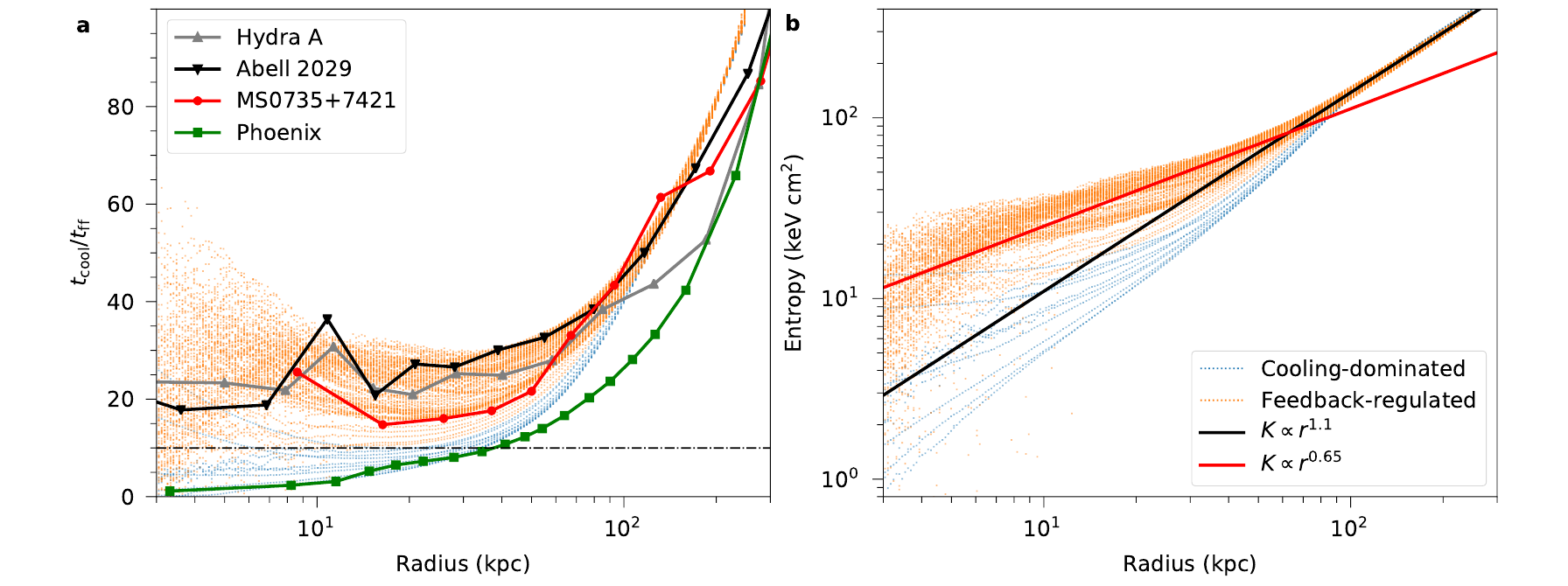}
\caption{\small {\bf Comparison of properties of the simulated X-ray emitting plasma with observations.} 
{\bf a}, The cooling-to-freefall timescale ratio derived from {the parent 10\,Gyr simulation, sampled every $\sim100$\,Myr} (blue and orange dotted lines), overlaid with profiles observed in four representative giant elliptical galaxies and clusters\cite{McNamara2016,Hogan2017,McDonald2019}. The timescale ratio profiles in the simulated cluster core vary considerably depending on whether they correspond to the period in time when radiative cooling dominates over AGN heating ($t<1$\,Gyr, blue dotted lines) or afterwards (orange dotted lines). With the exception of the Phoenix cluster, which maps into the cooling-dominated phase in our simulations, other systems exhibit a minimum ratio about 20. 
{\bf b}, Evolution of the entropy profile of the ICM over 10\,Gyr, sampled every $\sim100\,$Myr. In the stages regulated by AGN feedback, the entropy of the ICM inside the central 50\,kpc assumes a profile with a shallower slope, consistent with observations of early-type galaxies, galaxy groups, and clusters ($K\propto r^{0.65\pm0.11}$)\cite{Panagoulia2014,Babyk2018a}. {The slopes of observed power-law entropy profiles in the cores (red) and outskirts (black) are overplotted for comparison.}
}
\label{fig:profile}
\end{figure}

AGN feedback also plays a pivotal role for the thermal structure of the X-ray emitting ICM. {In Fig.~\ref{fig:profile} we show its properties in the original, 10\,Gyr simulation\cite{Qiu2018}}. Fig.~\ref{fig:profile}a shows the cooling-to-freefall timescale ratio ($t_{\rm cool}/t_{\rm ff}$), a measure that is commonly used to assess the thermal stability of the X-ray emitting ICM in galaxy cluster simulations\cite{Gaspari2012a, Li2015, Beckmann2019}. The timescale ratio profiles in the simulated cluster core vary considerably depending on whether they correspond to the stage when radiative cooling dominates over AGN heating ($t<1$\,Gyr in the simulation, see Methods) or vice versa. With the exception of the Phoenix cluster, which may be suffering from an unbridled cooling flow\cite{McDonald2012a,McDonald2019} and maps to the cooling-dominated phase of our simulation, all other systems exhibit a minimum ratio of about 20, consistent with observations of most cluster cores and early type galaxies\cite{McNamara2016, Hogan2017}. Interestingly, Abell~2029 has $t_{\rm cool}/t_{\rm ff}$ profile similar to others but is devoid of both molecular gas and nebular emission. This can be explained by the absence of AGN-driven outflows, if they, rather than thermal instability, are the key for the formation of cold gas filaments.

Our simulation reproduces additional properties of the X-ray emitting ICM, such as the velocity dispersion (Methods) and entropy profile, $K=k_{\rm B}T\,n_e^{-2/3}$. Fig.~\ref{fig:profile}b illustrates the evolution the entropy profile, which at $r\gtrsim 100$\,kpc asymptotes to the power-law, $K\propto r^{1.1}$, as expected for virialized ICM\cite{Voit2005}. In the evolutionary stages regulated by AGN feedback, the ICM at $r\lesssim 50$\,kpc assumes a profile with a shallower slope, consistent with those determined from observations of early-type galaxies, galaxy groups, and clusters ($K\propto r^{0.65\pm0.11}$)\cite{Panagoulia2014,Babyk2018a}. This agreement lands further support for the hypothesis that AGN feedback is a common driver of the properties of the ICM and the cold gas filaments. Ideas advanced in this work therefore provide new insights into a long-standing open question about the origin of the cold gas filaments and a new framework in which to interpret the impact of multiphase outflows driven by AGN feedback in galaxy clusters.

\begin{addendum}
 \item {Y.Q. thanks L.~C. Ho and S.~M. Faber for useful discussions. Y.Q. acknowledges support from the National Key R\&D Program of China (2016YFA0400702), the National Science Foundation of China (11721303, 11991052), and the High-performance Computing Platform of Peking University.} T.B. thanks the Kavli Institute for Theoretical Physics, where one portion of this work was completed, for its hospitality. Support for the early development of this work was provided by the National Aeronautics and Space Administration through Chandra Award Number TM7-18008X issued by the Chandra X-ray Center, which is operated by the Smithsonian Astrophysical Observatory for and on behalf of the National Aeronautics Space Administration under contract NAS803060. This research was supported in part by the National Science Foundation under Grant No. NSF PHY-1748958, and in part through research cyberinfrastructure resources and services provided by the Partnership for an Advanced Computing Environment (PACE) at the Georgia Institute of Technology, Atlanta, Georgia, USA.

 \item[Author contributions] Y.Q. performed the simulation and analyzed the data. Y.Q. and T.B. wrote the manuscript. Y.L., M.M., and B.R.M. commented on the manuscript, data analysis, and made suggestions for comparisons with observation. M.M. and B.R.M provided the observational data in Fig.\ref{fig:profile}a.
 
 \item[Competing interests] The authors declare no competing interests.

 \item[Correspondence and requests for materials] should be addressed to Y.Q. (yuqiu@pku.edu.cn).
 
\end{addendum}

\begin{methods}

\subsection{High cadence simulation of AGN feedback in galaxy clusters.} 
In order to follow the formation and evolution of the cold gas filaments on the fly, we re-simulated a 20\,Myr segment of a longer (10\,Gyr) galaxy cluster simulation that follows the evolution of a CCC in the presence of radiative and kinetic AGN feedback (simulation RT02 from our published simulation suite\cite{Qiu2018}). The new simulation is a ``zoom-in" in time, with high cadence of output set to about 50\,kyr. This 
frequency of output allows us to resolve the cooling and free-fall timescales for the multiphase gas
\begin{equation}
t_{\rm cool} = \frac{nk_{\rm B}TV}{(\gamma-1)L} \;\;\; {\rm and} \;\;\;  t_{\rm ff}=\sqrt{\frac{2r}{g(r)}},
\end{equation}
respectively. Here, $n$ is the number density of the gas, $L$ and $V$ are its total radiative luminosity and volume, respectively, $\gamma=5/3$ is the adiabatic constant, and $g(r)$ is the gravitational acceleration experienced by the gas as a function of the clustercentric radius. For example, $t_{\rm cool}\lesssim 10^7{\rm yr}$ for the gas cooling from $T=10^7\,$K to $10^4\,$K and $t_{\rm ff}\sim 10^8\,$yr for a clump falling from $r=30\,$kpc in our simulated cluster (see Fig.~\ref{fig:model}b). The starting point of the zoom-in simulation is chosen to capture the evolution of the network of cold filaments that forms around the 620\,Myr time mark of the parent simulation. The details of the numerical setup of the parent simulation also apply to the zoom-in simulation. We direct the reader to the publication\cite{Qiu2018} describing our simulation suite for more information and only summarize the main aspects below. 

The radiation-hydrodynamic simulation is performed using the adaptive mesh refinement code \texttt{Enzo}\cite{Bryan2014} with the ray-tracing radiative transfer package \texttt{Moray}\cite{Wise2011}. We model an isolated galaxy cluster experiencing AGN feedback, powered by accretion onto the central SMBH. The underlying potential of the cluster accounts for the contribution from the central SMBH\cite{Wilman2005}, the brightest central galaxy (BCG)\cite{Mathews2006}, and an NFW dark matter halo\cite{Navarro1996}, as detailed in Appendix~A of Qiu et al. (2019)\cite{Qiu2018}. The initial profiles for the density and temperature of the ICM are based on observations of the Perseus cluster\cite{Churazov2004, Mathews2006}. Simulations do not capture the self-gravity of the ICM. The accretion rate onto the central SMBH is estimated based on the gas properties in the vicinity of the black hole and is dominated by accretion of cold gas\cite{Tremblay2016}. We assume that 10\% of the accreted rest mass energy is channeled into the radiative and kinetic feedback from the AGN. The allocation of power between the two feedback modes is a function of the accretion rate of the SMBH in such a way that this setup produces a jet-dominated, radio-loud AGN at low accretion rates, and a radio-loud quasar at high accretion rates. To model the kinetic feedback, we accelerate the gas within 1\,kpc of the SMBH along a fixed jet axis (which corresponds to the vertical direction in figures shown in this paper) and into bipolar outflows. We model the radiative feedback as isotropic ionizing radiation from the central AGN using the ray-tracing module \texttt{Moray}. \texttt{Moray} calculates photoionization and Compton scattering leading to heating of the electrons along the path of each individual light ray. 

The cluster simulated in the parent simulation cools passively in the first 0.3\,Gyr, before AGN feedback is triggered by the formation and subsequent accretion of cold gas in the vicinity of the SMBH. The heating provided by the AGN feedback restores the thermal support of the cluster core within the first 1\,Gyr of this simulation. This initial phase of evolution corresponds to the blue lines in Fig.~\ref{fig:profile}. After the first 1\,Gyr, the AGN feedback regulates the evolution of the cluster core and its total luminosity is in the range $10^{43-46}\,{\rm erg\,s}^{-1}$.


\begin{figure}[t!]
\centering
\includegraphics[width=\linewidth]{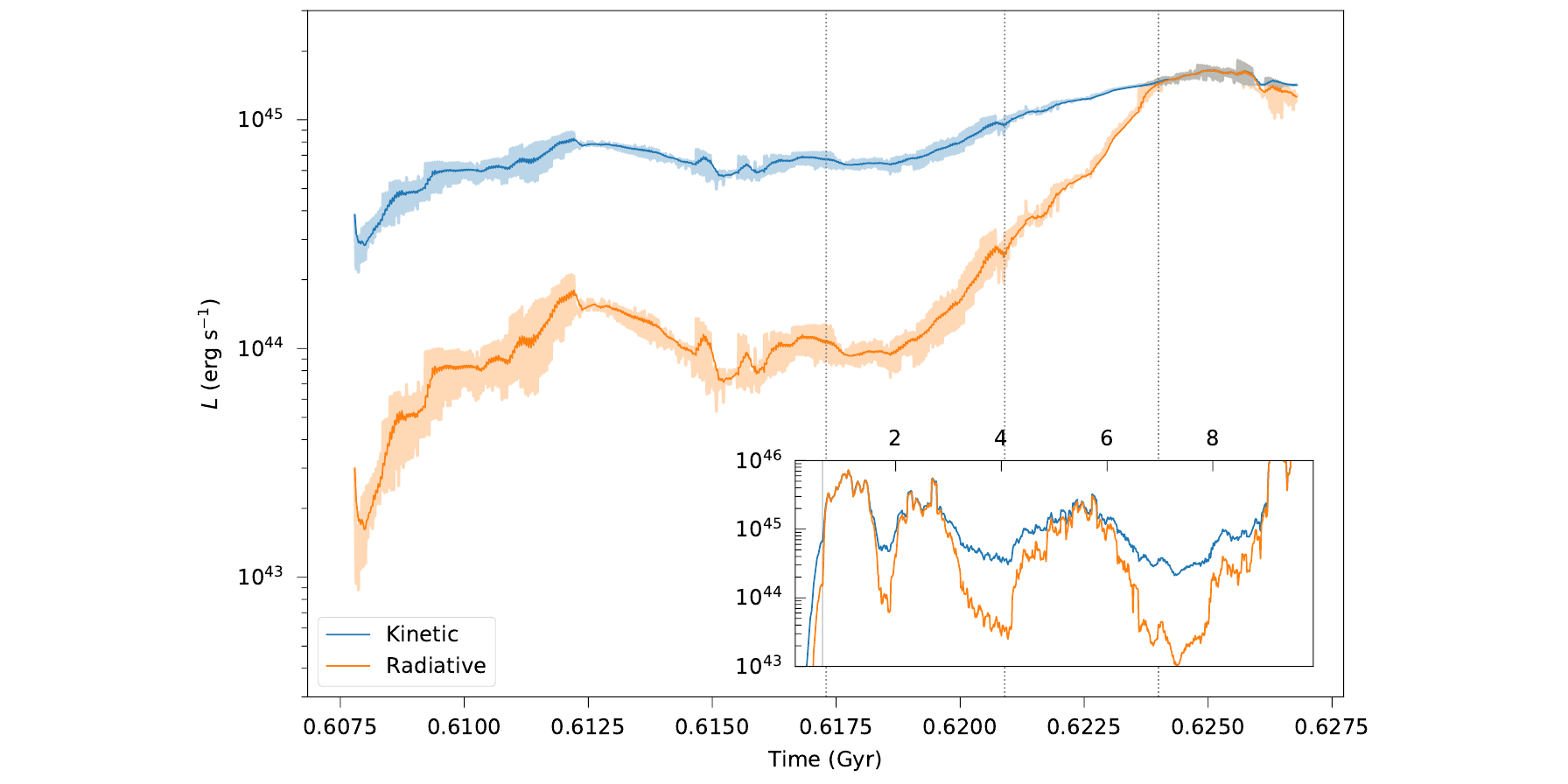}
\caption{\small {\bf Evolution of AGN feedback luminosity in the zoom-in simulation.} The total AGN luminosity is allocated to kinetic (blue) and radiative (orange) luminosity as a function of SMBH accretion rate. Lighter coloured bands correspond to the instantaneous AGN luminosities and darker lines show luminosities smoothed over 0.2\,Myr. Dotted vertical lines mark the instances in time corresponding to the panels in Fig.~\ref{fig:pic}a-c. Inset shows the evolution of the AGN luminosity over 10\,Gyr in the parent simulation, smoothed over $\sim0.1$\,Gyr. The grey vertical strip marks the duration of the high-cadence re-simulation.}
\label{fig:evo}
\end{figure}

\subsection{The importance of radiative and kinetic feedback for formation of warm outflows.} A phenomenon that is essential for formation of cold filaments in our simulation is the radiative feedback by the central AGN. Historically, the impact of radiative feedback on the evolution of the ICM has not been widely considered, due to its low coupling efficiency with the X-ray emitting plasma. This question became pertinent more recently, after observations established that CCCs commonly contain large amounts of cold atomic and molecular gas in their centres\cite{Hu1985, Johnstone1987, Heckman1989, Crawford1999, Jaffe2001, Edge2001, Edge2003, Salome2003a, Jaffe2005, Johnstone2007, Edwards2007, Hatch2007, McDonald2010, Oonk2010, Russell2019}. {In our simulation, the radiative feedback is responsible for producing the multiphase, ionized gas in the central $\sim 2$\,kpc. This gas is lifted by the outflows, driven predominantly by the kinetic, and to a lesser degree, radiative feedback. Because the presence of multiphase outflows is a prerequisite for the formation of the extended cold filaments, radiative feedback, in addition to the kinetic, is a necessary ingredient of any model that aims to study them.}

One important characteristic of radiative feedback in our simulations is that the luminosity of radiative feedback varies by a few orders of magnitude on relatively short timescales, set by the cooling time of the plasma in the cluster center. {This is illustrated in Fig.~\ref{fig:evo}, which shows the evolution of the kinetic and radiative luminosity over 20\,Myr in our zoom-in simulation. During this period of time, the radiative luminosity evolves from $\sim10^{43}$ to $\sim10^{45}$\,erg\,s$^{-1}$. In the simulation, every subsequent burst of AGN radiative feedback briefly photoionizes the cold gas. The photoionized gas quickly recombines, provides fuel for the SMBH, and in such a way triggers the new feedback outburst. The burst of $L_{\rm R}\sim10^{44}$\,erg\,s$^{-1}$ around $t=610$\,Myr, for example, produced the warm outflows that form the cold gas filament highlighted in the dashed box of Fig.~\ref{fig:pic}a-c (see also Supplementary Video~1).} The inset of Fig.~\ref{fig:evo} shows the long-term evolution of the kinetic and radiative luminosity over 10\,Gyr in the parent simulation. On both the shorter and longer timescales the kinetic luminosity (which has different dependence on the SMBH accretion rate in our feedback model) remains relatively level, and within a factor of few from $10^{45}$\,erg\,s$^{-1}$. In comparison, radiative feedback exhibits larger variability and a low duty cycle relative to the kinetic feedback. This property of our simulated AGN is consistent with observation that most BCGs in CCCs at low redshift do not host quasars.

Assuming that real AGNs in CCCs operate in the same way, the indirect evidence for the action of radiative feedback can be found in the presence of longer-lived extended cold filaments. Because the rise and freefall time of the filaments can be long ($\sim 10^8\,$yr at $r=30\,$kpc), they can point to a past outburst, during which the BCG briefly hosted a luminous AGN or a radio-loud quasar. A cluster which may be an illustration of this brief phase of evolution and presently hosts an X-ray luminous AGN ($L_{\rm R}\approx 5.6\times10^{45}\,{\rm erg\,s}^{-1}$) is Phoenix\cite{McDonald2015}. On the other hand, the Perseus cluster hosts a rich network of H$\alpha$ filaments but lacks an X-ray luminous AGN ($L_{\rm R}\sim10^{43}\,{\rm erg\,s}^{-1}$, assuming isotropic emission\cite{Fabian2015a}). We therefore speculate that the central AGN in Perseus was more luminous $\sim10^7-10^8$\,yr ago, much like the one in Phoenix, giving rise to the spectacular network of filaments. 

\subsection{The impact of ram pressure and radiative cooling on AGN-driven outflows.} 
In this section we describe the analytic model used to illustrate the impact of ram pressure and radiative cooling on multiphase outflows shown in Fig.~\ref{fig:model}b-c. In this model, spherical clumps of gas with mass $M=10^8\,M_\odot$ and temperatures $T=10^6$, $10^7$ and $10^8\,$K are launched from an altitude of $r=1\,$kpc in the stratified atmosphere described by the underlying cluster potential, and the initial temperature and density profiles of the simulated ICM. The cooling rate of the clumps due to the emission of radiation, $\Lambda(n,T)$, is calculated using the tabulated cooling rate for the gas in ionization equilibrium\cite{Sarazin1987} with metallicity $Z=0.01$, and is valid between $10^4-10^8$\,K. The clumps are assumed to be in pressure equilibrium with the surrounding ICM so that 
\begin{equation}
P(r,t)=P_{\rm ICM}(r)=\frac{\rho_{\rm ICM}(r)\,k_{\rm B}T_{\rm ICM}(r)}{{\mu m_p}} \,,
\end{equation}
where $\rho_{\rm ICM}(r)$ and $T_{\rm ICM}(r)$ are the density and temperature profiles of the ICM, $\mu=0.6$ is the mean atomic weight for fully ionized $Z=0.01$ plasma, and $m_p$ is the proton mass. This assumption allows us to calculate the density evolution of the spherical clump as
\begin{equation}
\rho(r,t)=\frac{\mu m_p\, P(r,t)}{k_{\rm B} T(r,t)}\,,
\end{equation}
and its instantaneous radius as $R(r,t)=(3V(r,t)/4\pi)^{1/3}$, where $V(r,t)=M/\rho(r,t)$ is the clump volume.

In addition to the gravitational acceleration, the gas clump traveling on a radial trajectory also experiences ram pressure from the surrounding ICM, $P_{\rm ram}(r,t)=\rho_{\rm ICM}(r)\,v_r(r,t)^2$, which modifies the acceleration of the gas clump as 
\begin{equation}
a(r,t)= g(r) \mp  P_{\rm ram}(r,t)\, \pi R(r,t)^2/M \,,
\end{equation}
where $g(r)<0$ is the gravitational acceleration, calculated from the underlying potential, and the minus (plus) sign corresponds to the clump traveling radially outward (inward). The evolution of each clump is then integrated with the time step $\Delta t=1$\,kyr, or 10\% of the cooling time, whichever is smaller. In each step, the temperature, density, volume, and radius of the clump are updated, allowing us to calculate the new acceleration, radial velocity, and position. 

The simplifying assumptions we make in this model include $(i)$ the spherical geometry of the gas clump, $(ii)$ mass loss due to ram pressure stripping of the clump is neglected, $(iii)$ apart from radiative cooling, energy exchange with the ambient medium is neglected, and $(iv)$ pressure equilibrium with the surrounding ICM is enforced in every step. More realistically, the geometry of the cold gas is filamentary and non-spherical, which further reduces the effect of ram pressure, and may allow the cold filaments in our model to travel even further. If similar (approximately cylindrical) geometry can be assumed for the warm and hot outflows, this may result in warm/hot outflows reaching further beyond the core region\cite{Kirkpatrick2015}. Ram pressure stripping reduces the cross-section ($A\propto R^2$) and the total mass of the clump ($M\propto R^3$) at the same time. Since deceleration due to ram pressure scales as $\propto A/M \propto R^{-1}$, it follows that the stripping of the clump also reduces the distance that it can travel, assuming that stripping results in smaller clumps. Moreover, the outflows can dissipate their kinetic energy in weak shocks that form at the interface of the outflows and the ICM. If so, some fraction of this kinetic energy can be converted to thermal energy, thus suppressing the reach of the gas clumps in the ICM atmosphere. Finally, given that the cluster we modeled on Perseus is roughly isobaric in the inner 10\,kpc, the assumption of pressure equilibrium does not have a significant impact on their trajectory in this region. Outside of this region, the ICM pressure gradually decreases, which in our model leads to a small expansion of the clump size, but this does not significantly enhance the influence of ram pressure. 

In addition to the kinematic properties of cold clumps, determined by the gravitational deceleration and ram pressure from the ICM, their evolution is also affected by their longevity. Because the clumps are typically moving through the ICM with velocities much higher than their sound speed ($c_{\rm s}\sim10$\,km\,s$^{-1}$), the shock heating can destabilize and dissipate the cold gas. The timescale for destruction of the cold clump due to shock heating can be estimated as the time for the shock to sweep over it\cite{Zhang2017}:
\begin{equation}
t_{\rm d} \sim \frac{R}{v_r} \left(\frac{\rho}{\rho_{\rm ICM}}\right)^{1/2}.
\label{eq:t_d}
\end{equation}
For a cold gas clump with radius $R\sim0.1$\,kpc in pressure equilibrium with the ICM, moving at $100-1000$\,km\,s$^{-1}$, the density contrast is $\rho/\rho_{\rm ICM} \sim 4,000$ and the resulting destruction timescale $t_{\rm d}\sim10^{7-8}$\,yr. We mark this timescale in Fig.~\ref{fig:model}bc, where the trajectory of the cold clump is shown as a dotted line after $10^7$ ($10^8$) years for the faster (slower) clump. This implies that most of the extended cold gas filaments are shredded in the ICM before falling back to the cluster center.



\begin{figure}[t!]
\centering
\includegraphics[width=\linewidth]{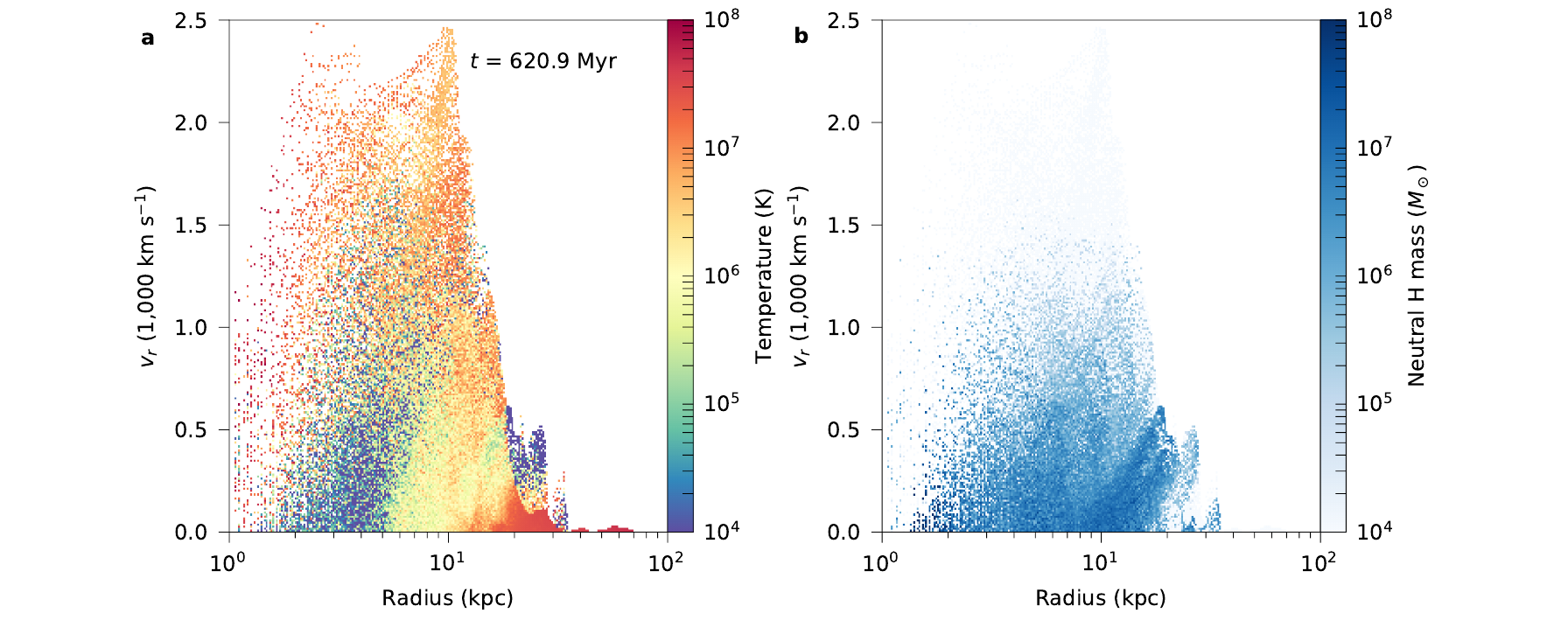}
\caption{\small {\bf Instantaneous radial velocity distribution for outflows in the high-cadence simulation.} 
Outflow speed as a function of radius, color-coded in terms of the mass-weighted temperature ({\bf a}) and neutral hydrogen mass ({\bf b}). The bulk of the cold gas is traveling with speeds below 1,000\,km\,s$^{-1}$, in agreement with observations and with predictions of the analytic model. The time of this snapshot corresponds to Fig.~\ref{fig:pic}b.}
\label{fig:phase}
\end{figure}

To assess how well our simple analytic model captures the salient points of the filament evolution, we compare its predictions with the radial velocity distribution for the multiphase outflows in our simulation. Fig.~\ref{fig:phase}a shows the instantaneous outflow speed as a function of the clustercentric radius for a snapshot in the simulation corresponding to Fig.~\ref{fig:pic}b. In the inner region of the cluster ($r\lesssim4\,$kpc), outflows are dominated by the hot ($T>10^7$\,K) and warm plasma ($10^4<T\leq10^7$\,K), which is launched at speeds $>1,000\,{\rm km\,s^{-1}}$. The warm phase of the outflow reaches $\sim10$\,kpc, while still at relatively high speeds, after which point it decelerates and cools rapidly, giving rise to the lower speed cold outflows. This is illustrated by the dramatic drop of the outflow velocity at $r\sim10$\,kpc, similar to the evolution of the $T=10^7$\,K gas clump in Fig.~\ref{fig:model}c. Once the clumps cool below $T\sim10^4\,$K, they experience negligible impact of the ram pressure on their kinematics, which enables them to travel out to $\sim30$\,kpc, in agreement with the prediction of our model. It is worth noting that the $10^6$\,K gas clump cools to $10^4\,$K in $\lesssim 10^5\,$yr and hence appears as a blue/purple line throughout its evolution in Fig.~\ref{fig:model}bc.

Because modern observatories are incapable of measuring the outflow velocities of hot and warm outflows directly, in practice, the best constraints on kinematics of outflows come from the velocity measurements of the cold gas. Fig.~\ref{fig:phase}b shows the radial velocity distribution of the cold gas only (traced by neutral hydrogen) in the same simulation snapshot. The bulk of the cold gas is traveling with speeds below 1,000\,km\,s$^{-1}$, in agreement with the expectation of our analytic model for $10^7$\,K gas clump, illustrated in Fig.~\ref{fig:model}c. {The kinematics of the $10^6$\,K gas clump shown by the analytic model is not reproduced in the simulation, which is devoid of cold outflows with speeds $\sim 2,000 - 3,000\,{\rm km\,s^{-1}}$. As discussed earlier, because of their short cooling timescale, the clumps with temperature of $10^6$\,K or less are difficult to accelerate to high velocities before they become compact and immune to the effects of ram pressure. This comparison of the analytic model and simulation leads us to conclude that the gas phase launched in AGN outflows that ultimately gives rise to the most extended cold filaments resides in the temperature range $10^6\lesssim T\lesssim 10^7$\,K.} This is of interest because simulated velocity distribution shown in Fig.~\ref{fig:phase}b is broadly consistent with observations of the cold gas filaments in CCCs\cite{McDonald2012}, which find that the H$\alpha$ emission lines associated with the filaments have widths (FWHM) of $100-600$\,km\,s$^{-1}$. Similar results have been recently obtained from ALMA observations of BCGs\cite{Russell2019}, which show that molecular gas, found between a few to tens of kpc from the cluster centre, has velocities around a few hundred km\,s$^{-1}$, much lower than the corresponding free-fall speed. These findings do not favor a simple ballistic motion as an explanation for the kinematics of the cold gas. The combination of multiphase outflows, radiative cooling, and ram pressure therefore appears to be the key in explaining this phenomenon. Furthermore, other physical processes that are not captured by our simulation, such as magnetic stresses in the ICM, may also help to reduce the outflow speeds.

\subsection{The evolution and dust properties of the AGN-driven outflows.} 
%

\begin{figure}[t!]
\centering
\includegraphics[width=\linewidth]{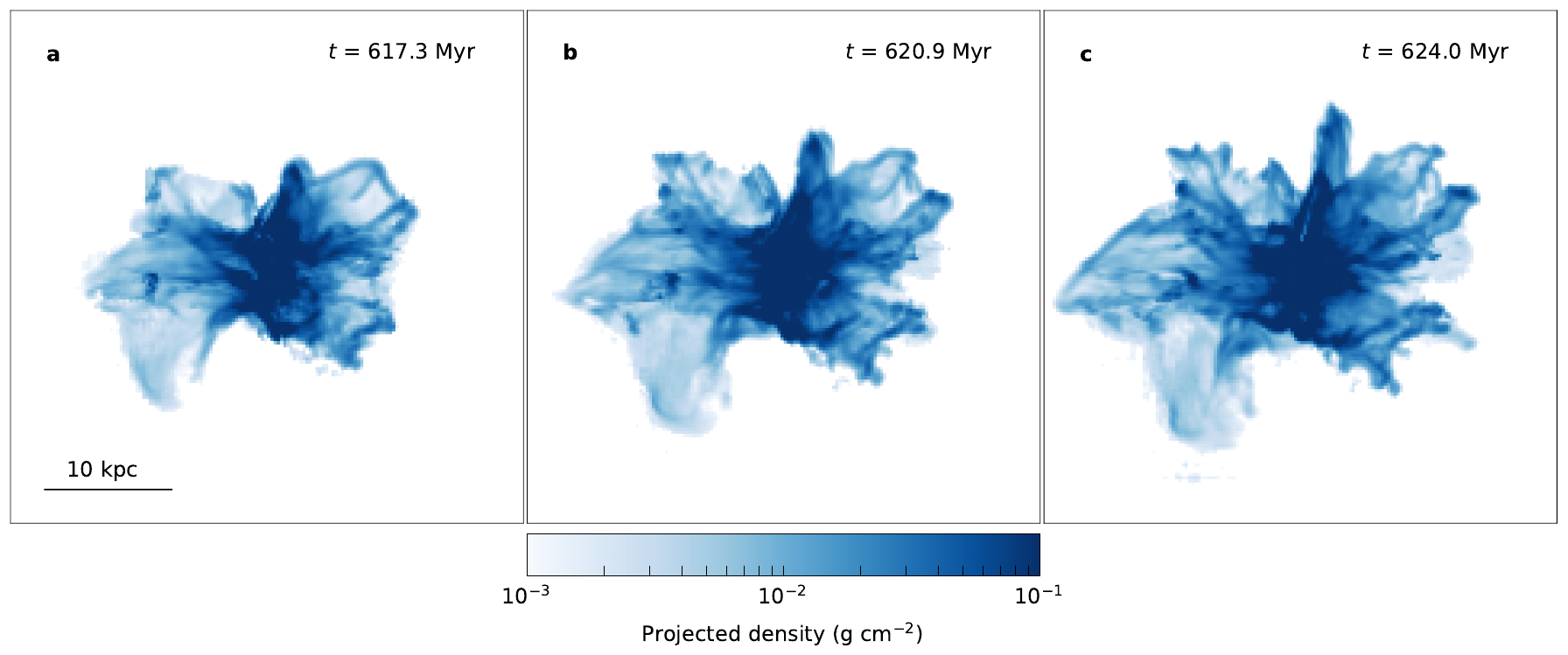}
\caption{\small {\bf Origin of the cold filaments.} 
{\bf a-c}, Projected gas density showing the spatial reach of the AGN-driven material, traced by a passive fluid injected from the central 1\,kpc, corresponding in time to Fig.~\ref{fig:pic}a-c. A movie showing the evolution in the high-cadence simulation is provided in Supplementary Video 1.}
\label{fig:tracer}
\end{figure}

{In order to understand the origin of the cold filaments, we inject a passive tracer fluid in the central 1\,kpc of our high-cadence simulation and follow the subsequent evolution. Panels in Fig.~\ref{fig:tracer} show the projected gas density of computational cells that contain the tracer fluid at a given time in the simulation. The three columns correspond in time to those in Fig.~\ref{fig:pic}a-c, and show the spatial reach of the outflows launched after $t = 608$\,Myr, the start of this re-simulation. After about 15\,Myr the gas that originated in the inner 1\,kpc is dispersed under the influence of AGN feedback and reaches several tens of kiloparsecs. This illustrates the role of AGN feedback in transporting metals and dust from the central galaxy to the ICM. A visual comparison of these maps with cold filaments in Fig.~\ref{fig:pic}a-c, indicates that the filaments contain gas that was launched from the central 1\,kpc. Note however that outflows launched from the central 1\,kpc before the injection of the tracer fluid cannot be shown by this method.}
 
{If the multi-phase outflows from the central kpc carry any amount of dust, the dust grains will be eroded by thermal dust sputtering\cite{Draine1979,Li2019a} at the rate given by\cite{Draine2011}
\begin{equation}
\frac{da}{dt}\approx-\frac{1\times10^{-6}}{1+T_6^{-3}}\left(\frac{n_{\rm H}}{{\rm cm}^{-3}}\right)\,{\rm \mu m\,yr}^{-1},
\label{eq_sputter}
\end{equation}
where $a$ is the size of the dust grain, and $T_6\equiv T/ 10^6\,{\rm K}$ and $n_{\rm H}$ are the temperature and hydrogen number density of the surrounding warm gas, respectively. In order to explicitly determine the temperature threshold beyond which no dust grains are expected to exist in outflowing gas, we use the analytic model presented earlier and calculate the sputtering rate given by equation~\ref{eq_sputter} along with the radiative cooling along the clump's trajectory. For a given initial temperature of the clump, we calculate the smallest size of a grain that may survive the erosion and show it in Fig.~\ref{fig:dust}a. We find that in the clumps with the temperature higher than $4\times10^6$\,K the sputtering depth exceeds 1\,${\rm \mu m}$. This implies that only outflows with the initial temperature below this threshold have a chance of preserving the dust drawn from the central 1\,kpc of the cluster.}


\begin{figure}[t!]
\centering
\includegraphics[width=\linewidth]{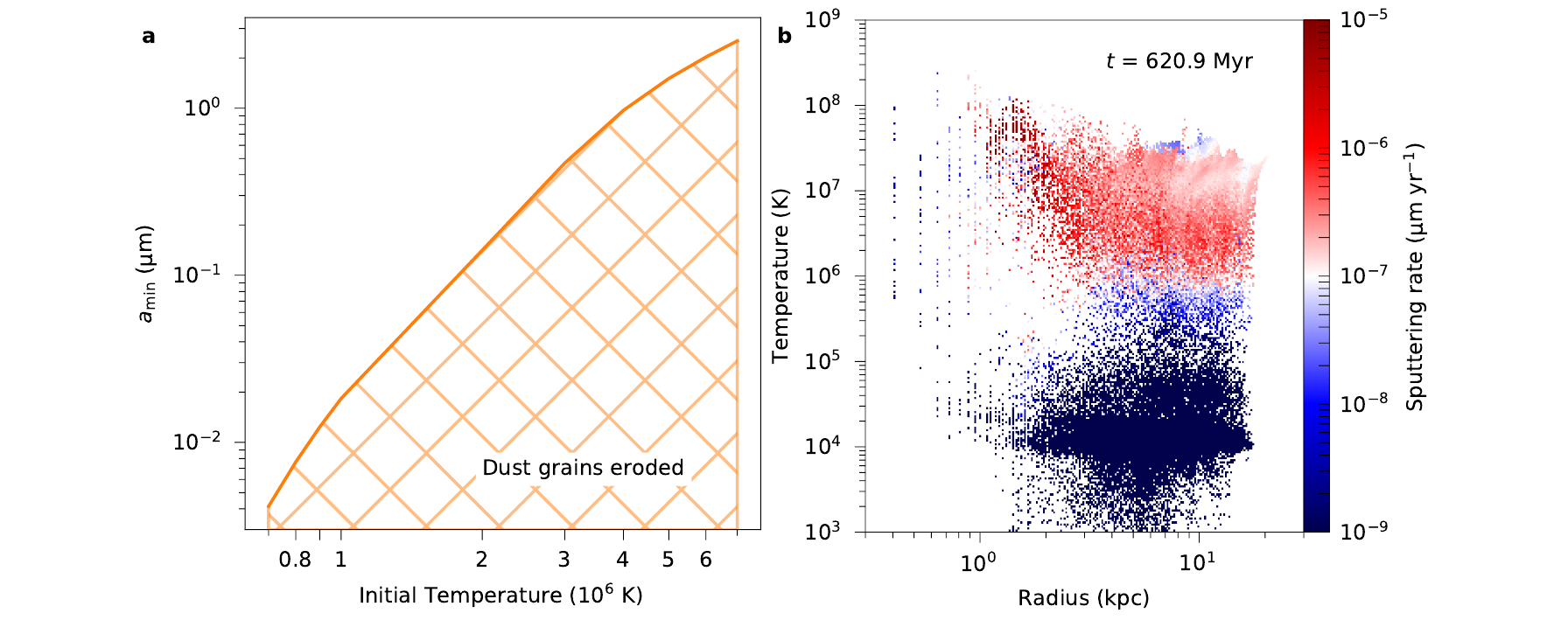}
\caption{\small {\bf Dust erosion in warm outflows.} 
{\bf a}, The size of a smallest dust grain that can survive in the warm outflows as a function of their initial temperature. For characteristic dust grain size of $1\,{\rm \mu m}$ or less, most dust content is eliminated in outflows with initial temperature higher than $4\times10^6$\,K.
{\bf b}, Radial temperature distribution of the tracer fluid, color-coded with the thermal dust sputtering rate, weighted by cell mass. For dust size $a\sim 0.1\,{\rm \mu m}$, blue colour corresponds to a sputtering timescale of $t_{\rm sp} \gtrsim$1\,Myr. This is longer than the cooling time of the warm outflows, indicating that dust can be preserved in the outflowing gas.}
\label{fig:dust}
\end{figure}

{Fig.~\ref{fig:dust}b shows the radial temperature distribution of the tracer fluid, colour-coded with the thermal dust sputtering rate and weighted by cell mass. Less than 10\,Myr after the beginning of the simulation the presence of the tracer fluid is evident in all components of the multi-phase ICM.
This panel indicates that the hot phase of ICM (traced mainly by red color) has the highest sputtering rate. The warm and cold gas phase (blue color) are characterized by lower sputtering rates that correspond to a sputtering timescale, $t_{\rm sp} \approx a/(da/dt) \gtrsim 1$\,Myr, for a grain size $a\sim 0.1\,{\rm \mu m}$. The sputtering timescale is therefore shorter than the timescale for outflows to reach tens of kpc ($t_{\rm rise} \sim 10^{7-8}$\,yr) but is comparable to or longer than the cooling timescale of outflows with initial temperature $\sim {\rm few} \times 10^6$\,K (see Fig~\ref{fig:model}). The hierarchy of timescales, $t_{\rm cool} \lesssim t_{\rm sp} < t_{\rm rise}$, indicates that the warm outflows are expected to cool before their dust content is depleted and that this evolution takes place while they are still rising. We therefore conclude that the dust can be preserved in the warm outflows launched by the AGN from the inner kpc.}

{Observationally, the amount of dust reported for the Perseus cluster is $\sim 10^7\,M_\odot$\cite{Mittal2012}. Given the total amount of $\sim 10^{11}\,M_\odot$ observed cold gas in the filaments\cite{Fabian2011b}, this implies a dust-to-gas ratio of only $\sim10^{-4}$, two orders of magnitude lower than the galactic value. To estimate the dust content in our simulated filaments, we must also consider the dilution due to mixing with the surrounding ICM. In order to investigate the degree of mixing between the ICM and the outflows, we examined the components of the cold filaments at the end of the simulation. Using the tracer fluid, we find on average $\sim20$\% of the filament mass originates from outflows launched from the central 1\,kpc, in filaments that are rising with velocities larger than $300\,{\rm km\,s}^{-1}$. This fraction decreases to $\sim5$\% when we do not include the above velocity constraint (and therefore include the lower velocity and infalling cold filaments). This suggests that as the outflows travel in the cluster, they gradually mix with the ICM. Therefore, the rising filaments closer to the AGN have higher dust-to-gas ratio, compared with the filaments that have traveled to larger radial distances.

A similar trend is observed in NGC 1275, the system we model. Spitzer data\cite{Johnstone2007} showed that filaments closer to the cluster center ($r\sim10$\,kpc) have detectable 11.3-${\rm \mu m}$ polycyclic aromatic hydrocarbon (PAH) feature in the spectra, while the filaments in two other regions farther away ($r\sim20$\,kpc) lack this feature. The radial distribution of PAH in NGC 1275 can be interpreted as the result of mixing and thermal sputtering, where $r\sim10$\,kpc is roughly the threshold at which large dust grains ($a\gtrsim 0.1$\,${\rm \mu m}$) are eroded to the smallest detectable size, and are either absent or undetectable beyond this radius. This radial trend in dust content is consistent with predictions based on our simulation. In comparison, the filamentary system in NGC 4696 has spatial extent $r<5$\,kpc and enough dust to form visible dust lanes co-spatial with the filaments\cite{Fabian2016b}. {With a $\sim 10^{-3}$ dust-to-gas ratio\cite{Mittal2011}}, this implies that in systems such as NGC 4696 with less extended filaments, evolution of the outflows may allow for higher dust-to-gas ratio, due to a smaller degree of mixing with the ICM. 

Beyond thermal sputtering and mixing with the ICM, several additional processes not captured in our simulation may work to increase the dust content. We summarize them below:
(a) Dust growth in cold filaments. Once the outflow temperature drops below $10^5$\,K, accretion growth of the dust grains by capturing gas atoms becomes the dominant process over sputtering in the dust evolution\cite{Draine1979}, which helps to increase the dust-to-gas ratio. 
(b) Star formation. Dusty cold gas filaments provide natural conditions for star formation, which can further populate the surrounding gas with dust.
(c) Magnetic fields. It has been proposed that filaments are supported by magnetic pressure\cite{Fabian2008}, which may also help to shield the dusty outflows by deflecting ionized particles. 
(d) Dust formation in AGN winds. It was recently proposed that AGN wind provides suitable conditions for the formation of substantial amounts of dust\cite{Sarangi2019a}. This will further fortify the AGN-driven outflows with dust, in line with our model that the warm outflows are the key in forming dusty cold filaments in galaxy clusters.
}


\subsection{Comparison with observations and other models for cold filament formation.} 
The origin of the cold gas filaments in CCCs and their relationship with the central AGN remains an important open question in the context of galaxy clusters and the cooling flow problem. This question is part of a broader effort to understand how AGN feedback couples to the host galaxy and surrounding environment. It is closely related to recent investigations of active galaxies that show evidence for large-scale multiphase outflows, whose kinematic and ionization properties are stratified as a function of distance from the SMBH\cite{Tombesi2013,Tombesi2015}. These works propose that (a) highly ionized, relativistic outflows, launched from the sub-parsec size region centered on the SMBH and (b) warm X-ray absorbers and molecular outflows moving with speeds of $\sim 10^2-10^3\,{\rm km\,s^{-1}}$, actually represent parts of a single large-scale stratified outflow observed at different locations from the black hole. While CCCs provide notably different environments for AGN feedback, these studies provide support for a basic premise of this work: that observed AGNs, like the AGN in our simulation, can launch large-scale multiphase outflows. The new element of this work is that the cold filaments form out of the warm outflows, while they are rising and expanding in the cluster core. Driven by inertia, the filaments initially rise and expand until they reach the turning point of their trajectory and start falling toward the cluster center. This behavior is captured in Fig.~\ref{fig:pic} and Supplementary Video~1.

The existing models for the formation of cold filaments predict that they form out of thermally unstable ICM, in locations where its cooling time scale ($t_{\rm cool}$) falls below some multiple of the dynamical, free-fall timescale ($t_{\rm ff}$). The cold gas in this scenario is assumed to form in-situ, before raining down to the cluster centre, in a process also described as ``precipitation''\cite{Voit2015a}. {Besides the outflowing cold filaments described earlier, we also identify the infalling filaments in our simulation. {The infalling filaments largely originate from the outflows that are past the turning point of their trajectory and are falling toward the cluster center. We verify that the outflow and inflow mass rates of cold filaments are comparable, leaving limited room for contribution from an additional, precipitation driven component. Furthermore, we note that the contribution to infalling filaments from precipitation is also unlikely because $t_{\rm cool}/t_{\rm ff}$ measured in our simulation is higher than the precipitation limit in the stages regulated by AGN feedback, as shown in Fig.~\ref{fig:profile}a. We therefore conclude that even if precipitation is present in our simulation, it is subdominant and not necessary in order to produce cold filaments.}}

An important observational constraint that has been used as a criterion for cold gas formation in theoretical models is set by the cooling-to-freefall timescale ratio, $t_{\rm cool}/t_{\rm ff}$. This ratio is in observed systems primarily determined by their cooling time, as the freefall time profile is similar among ellipticals and central cluster galaxies. Namely, observations of most giant ellipticals and BCGs find the minimum value of this ratio to be about 20\cite{McNamara2016,Hogan2017}. The value of the ratio commonly inferred from the precipitation models is $<10$, in tension with observations of clusters that contain cold gas in their centres. Some studies argue $t_{\rm cool}/t_{\rm ff}=10$ is a floor which is breached only rarely, when systems are in a rapidly cooling state\cite{Voit2015a}. The incidence of systems lying below the timescale ratio of 10 is however much lower than that of systems with ongoing cooling into molecular clouds, which appear to be common. If so, the thermal instability of the ICM may not fully explain the existence of cold gas in real systems. {It is worth noting however that recent simulations of precipitation report evidence for cold gas formation at $t_{\rm cool}/t_{\rm ff}$ ratios as high as 20\cite{Prasad2018, Beckmann2019}, and are closer to reconciling the predictions of this mechanism with observations.} In comparison, thermal properties of the X-ray emitting plasma predicted by this work are in good agreement with that of observed systems in terms of both the timescale ratio and entropy (as shown in Fig.~\ref{fig:profile}). This agreement lands further support for the hypothesis that AGN feedback can be a common driver of the properties of the ICM and the cold gas filaments.


\begin{figure}[t!]
\centering
\includegraphics[width=\linewidth]{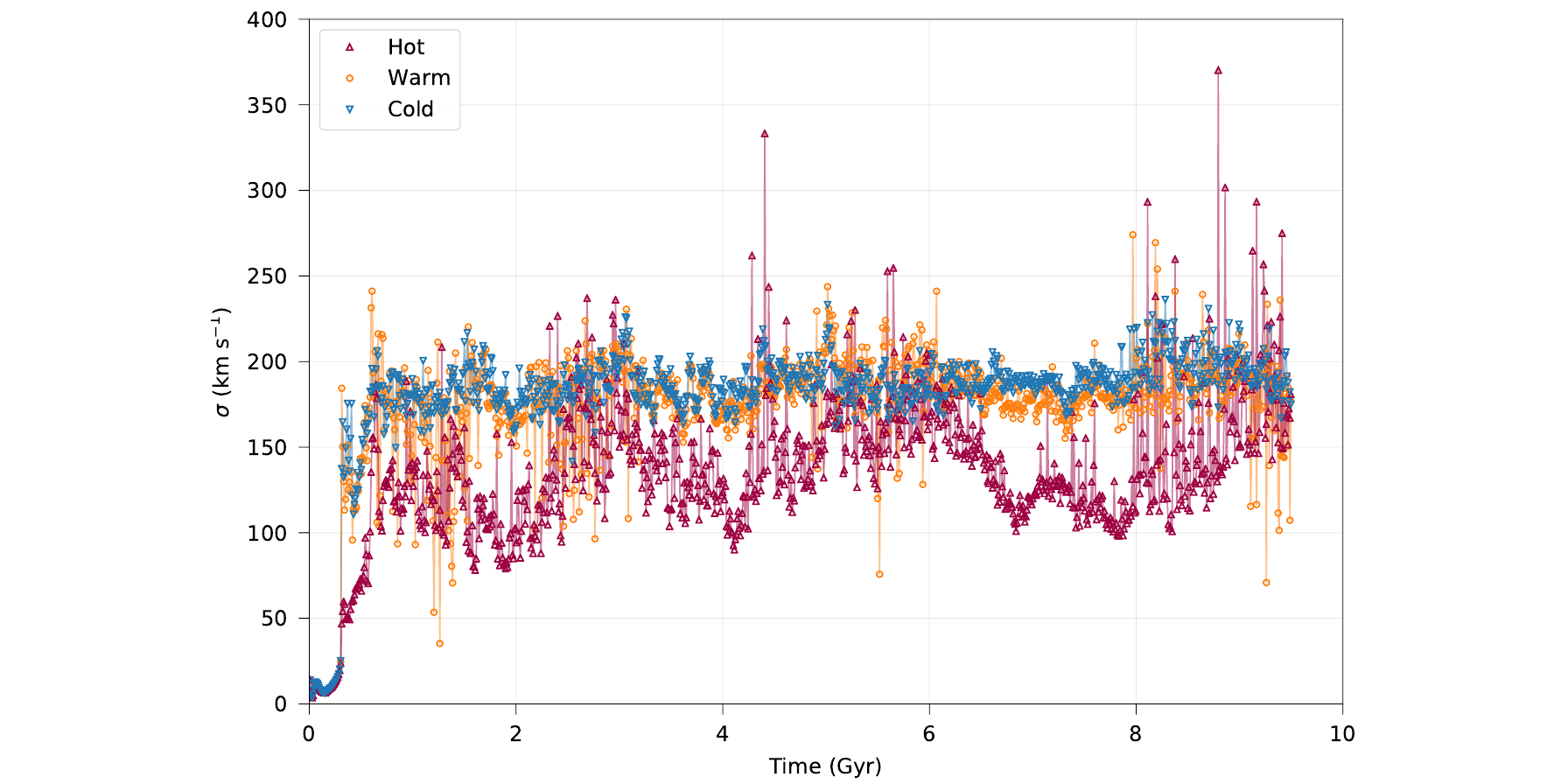}
\caption{\small {\bf Velocity dispersion of different temperature phases of the gas in simulated cluster core.} 
The resulting velocity dispersion is calculated as the average of the $x$, $y$, and $z$ components, each weighted by the X-ray emissivity (for hot and warm gas) and H$\alpha$ emissivity (for cold gas), at radii between 3 and 20\,kpc. While warm and cold gas share similar values, the velocity dispersion of the hot gas shows stronger, uncorrelated variability, indicating a separate kinematic origin from the cold gas filaments.}
\label{fig:sigma}
\end{figure}

We also examine the velocity dispersion of the gas, $\sigma$, used as a probe of its kinematic properties. Fig.~\ref{fig:sigma} shows the temporal evolution of $\sigma$ for gas in different temperature ranges in our simulation, at radii corresponding to the extent of cold gas filaments, $3\,{\rm kpc}<r<20\,{\rm kpc}$. The velocity dispersion value at a given time is calculated as the average of the $x$, $y$, and $z$ components. $\sigma$ of the hot and warm gas are weighted by their respective X-ray emissivities measured in the $0.1-10$\,keV range, while $\sigma$ of the cold gas is weighted by its H$\alpha$ emissivity arising from recombination\cite{Draine2011}. This approach allows us to reproduce more closely the properties of the observed data and to compare our $\sigma$ values with the X-ray and optical observations of cluster cores\cite{Hitomi2016,GM2018}. The trend worth pointing out in Fig.~\ref{fig:sigma} is that the warm and cold gas components share similar values, $\langle\sigma_{\rm warm}\rangle=179\pm25\,{\rm km\,s}^{-1}$ and $\langle\sigma_{\rm cold}\rangle=187\pm14\,{\rm km\,s}^{-1} $. Meanwhile, the velocity dispersion of the hot gas shows stronger variability with $\langle\sigma_{\rm hot}\rangle=140\pm36\,{\rm km\,s}^{-1} $, and has higher values during the peak of AGN activity. This indicates independent kinematic properties of the hot gas and the cold gas filaments, which is expected in the context of our work, since the hot outflows are not the driving force for the formation of the cold gas filaments. In contrast, some precipitation models predict correlation between the velocity dispersion of the hot and cold gas\cite{Gaspari2017a}, because in the alternative scenario the cold gas condenses out of the hot ICM and their kinematic evolution is related. Therefore, measurements of $\sigma$ for different gas phases in multiple CCCs can in principle be used to test the validity of a subset of models.

Another challenge associated with the class of models that describe thermal instability and precipitation of cold gas from the hot ICM is that filaments formed in this way should be devoid of any dust content, as dust is quickly destroyed in the hot plasma with temperature higher than a $\sim {\rm few}\times10^6$\,K\cite{Li2019a}. This prediction is in contrast with observations that unambiguously detect presence of dust in cold gas in CCCs\cite{Edge2010,Donahue2011}. {The scenario proposed in this paper, within which cold filaments mainly form from warm outflows with temperature $\lesssim10^7$\,K provide a natural solution to this puzzle, as they can preserve the dust content drawn from the cluster core (see also Fig.~\ref{fig:dust})}. 

An additional prediction of our model pertains to the relationship between the observable properties of the H$\alpha$ filaments and the kinetic and radiative feedback from SMBHs in BCGs\cite{Qiu2019}. This relationship was studied in detail on a full suite of 3D radiation-hydrodynamic simulations, including the parent simulation to the zoom-in simulation used in this work. {In contrast to some simulations supporting precipitation\cite{Beckmann2019}, we find that the AGN feedback in CCCs promotes the formation of spatially extended H$\alpha$ filaments and that their presence indicates an ongoing or a recent outburst of AGN feedback.} {A direct consequence of this causal relationship is that the mass and spatial extent of the cold filaments positively correlate with the luminosity of AGN feedback in our simulations\cite{Qiu2019}.} Both can therefore be used as independent probes of the AGN activity.

In another theoretical scenario that merits consideration, the cold filaments are entrained by the outflows and lifted from the central cluster galaxy, where they existed in the first place. This model faces several challenges. Firstly, it is difficult to accelerate cold gas through the hot ICM to high speeds without disrupting it\cite{Zhang2017} (see equation~\ref{eq:t_d}). Even if that were possible, one can show that the filaments that extend beyond 10\,kpc would have to be launched with initial velocities $>1,000\,{\rm km\,s}^{-1}$. Such high velocities are however not seen in observations of cold filaments in galaxy clusters\cite{McDonald2012, Russell2019}. This indicates that even if some initially cold gas is entrained in the outflows, it does not seem to attain speed $>1,000\,{\rm km\,s}^{-1}$, and should therefore be unable to reach altitudes of 10s of kpc. It is worth noting that the velocity distribution of the cold gas filaments produced in our simulation (and elucidated by our analytic model) is similar to that seen in observations of CCCs, without imposing special requirements on their geometry (see analysis presented in the previous section).

\subsection{Calculation of the X-ray images, entropy and cooling profiles of the X-ray emitting plasma.} 
In this section we describe the procedure used to calculate the synthetic X-ray images in Fig.~\ref{fig:pic}d-f as well as the entropy and cooling profiles of the X-ray emitting plasma shown in Fig.~\ref{fig:profile}. The first step in generating the X-ray images is to calculate the X-ray emission from the simulated, 3D snapshot of the cluster. We use the \texttt{yt}\cite{Turk2011} and \texttt{pyXSIM}\cite{ZuHone2016a} packages to generate the X-ray photons in the $1-7$\,keV energy band emitted in the central 200\,kpc of the simulated cluster. These energy thresholds are chosen so as to avoid the filamentary X-ray emission that appears at lower energies, and to reflect the energy range usually shown in observed images. The X-ray photons are generated assuming thermal emission of the collisional plasma with metallicity $Z=0.011$, based on the \texttt{APEC} code\cite{Smith2001}.

Subsequently, the cluster is placed at the redshift of Perseus, $z=0.0179$ (assuming ${\rm \Lambda CDM}$ cosmology, $h = 0.71$, $\Omega_m = 0.27$, $\Omega_\Lambda = 0.73$, this implies an angular scale of 0.359\,kpc per arcsec), and the photons are projected along the line of sight and processed through a T\"{u}bingen-Bolder foreground galactic absorption model\cite{Wilms2000}, assuming hydrogen absorption with a column density of $N_{\rm H}=4\times10^{20}$\,cm$^{-2}$. The 2D map of the X-ray photons calculated in this way is then convolved with the instrument response file using the \texttt{SOXS}\cite{ZuHone2018} package. The file is a simplified version of the Cycle-0 Chandra response matrix, with angular resolution of 0.5 arcsec, and with the chip gaps removed. The length of the exposure is set to 1\,Ms. In order to enhance the contrast of more subtle features in the ICM, such as the X-ray cavities, we calculate the fractional variance of this image by subtracting from it the azimuthally averaged profile, and subsequently dividing the difference by the azimuthal average. The image processing results in the cavities in the inner 3\,kpc, which are an artifact of the visualization that arises due to a large variation in X-ray surface brightness in this region. These regions are intensely heated by the AGN feedback and represent locations where bipolar outflows launched in the simulation collide with the surrounding ICM. This gives rise to an inner pair of cavities in Fig.~\ref{fig:pic}d-f. 

The radial entropy profile in Fig.~\ref{fig:profile}b, $K(r)$, is calculated as a spherical average in each radial shell of the simulated cluster. In the calculation of the average, the contribution from each cell in the shell is weighted by its X-ray luminosity in the energy range $0.1-10$\,keV. The X-ray luminosity used in this calculation is obtained using the pyXSIM package, as described above. It is worth noting that some fraction of the $T\lesssim10^7$\,K gas is co-spatial with the cold gas filaments (these are the so-called warm X-ray filaments\cite{Fabian2006}). In these regions the proximity of the cold gas and warm X-ray gas may lead to the non-radiative cooling of the X-ray gas through mixing with the cold gas. Similarly, the X-ray emission by the same plasma may be suppressed due to heavy absorption and scattering of the emitted X-ray photons by nearby dust. In order to take these effects into account and avoid an overestimate of the X-ray emission from such regions (since our simulations do not capture these effects), we exclude from the calculation of $K(r)$ the cells that are immediate neighbours with the cold gas cells. Following the convention common in X-ray observations, we use the electron number density, $n_e$, to calculate $K=k_{\rm B}T\,n_e^{-2/3}$, and note that this results in the value of entropy slightly higher than the value that we would obtain if we used the total plasma number density instead. 
The cooling-to-freefall timescale ratio profile in Fig.~\ref{fig:profile}b, $t_{\rm cool}(r)$, is calculated by summing over the thermal energy of the gas with $k_{\rm B}T>0.1$\,keV in each radial shell, and then dividing the sum by the $0.1-10$\,keV X-ray luminosity of the shell. For the same reason as mentioned above, the gas cells adjacent to the cells containing cold gas are not included in the calculation. 

\end{methods}

\noindent{\bf Data availability}
The data that support the plots within this paper and other findings of this study are available from the corresponding author upon reasonable request.

\noindent{\bf Code availability}
The simulations presented here were performed using an adapted version of the \texttt{Enzo} code. The main repository is available at http://enzo-project.org/, and the customizations made for this work are available from the corresponding author upon request.


\noindent{\bf References}
\small


\end{document}